\documentclass[11 pt]{article}
\usepackage{comment} 
\usepackage[a4paper, tmargin=.75in, lmargin=.75in, bmargin=1in, rmargin=.75in]{geometry}
\usepackage{graphicx}   
\usepackage{comment} 
\usepackage{amssymb,amsmath, mathrsfs,mathtools}
\usepackage[font=scriptsize,labelfont=bf]{caption}
\usepackage{setspace}
\usepackage{tabularray}
\usepackage{multirow}
\usepackage{array}
\usepackage{bigints}
\usepackage{xurl}
\usepackage{authblk}
\usepackage{nameref}
\usepackage{lineno}
\usepackage{soul}  
\usepackage[sort&compress,numbers]{natbib}
\usepackage[colorlinks,allcolors=blue]{hyperref}

\providecommand{\keywords}[1]
{\small \textbf{\textit{Keywords---}} #1}

\renewenvironment{abstract}{%
  \textbf\abstractname
  \list{}{\vspace{-0.2 cm} \leftmargin0in \rightmargin\leftmargin}
  \item\relax
}{%
  \endlist \par\bigskip
}

\usepackage{titlesec}
\titleformat*{\section}{\large\bfseries}
\titleformat*{\subsection}{\normalsize\bfseries}

\newcommand{\delete}[1]{}

\title{\Large{Real-time Estimation of Bound Water Concentration during Lyophilization with Temperature-based State Observers}}

\author[1,2,3]{\normalsize Prakitr Srisuma}
\author[1,3]{\normalsize George Barbastathis}
\author[1,2*]{\normalsize Richard D. Braatz}

\affil[1]{\scriptsize Center for Computational Science and Engineering, Massachusetts Institute of Technology, Cambridge, MA 02139, USA} 
\affil[2]{\scriptsize Department of Chemical Engineering, Massachusetts Institute of Technology, Cambridge, MA 02139, USA}
\affil[3]{\scriptsize Department of Mechanical Engineering, Massachusetts Institute of Technology, Cambridge, MA 02139, USA}

\date{}

\begin{document}
\maketitle
\def\thefootnote{*}\footnotetext{Corresponding author. Email: braatz@mit.edu}
\def\thefootnote{**}\footnotetext{\copyright \ 2024. This manuscript version is made available under the CC-BY-NC-ND 4.0 license \url{https://creativecommons.org/licenses/by-nc-nd/4.0/}.}
\hrule

\begin{abstract}
Lyophilization (aka freeze drying) has been shown to provide long-term stability for many crucial biotherapeutics, e.g., mRNA vaccines for COVID-19, allowing for higher storage temperature. The final stage of lyophilization, namely secondary drying, entails bound water removal via desorption, in which accurate prediction of bound water concentration is vital to ensuring the quality of the lyophilized product. This article proposes a novel technique for real-time estimation of the bound water concentration during secondary drying in lyophilization. A state observer is employed, which combines temperature measurement and mechanistic understanding of heat transfer and desorption kinetics, without requiring any online concentration measurement. Results from both simulations and experimental data show that the observer can accurately estimate the concentration of bound water in real time for all possible concentration levels, operating conditions, and measurement noise. This framework can also be applied for monitoring and control of the residual moisture in other desorption-related processes.

\end{abstract} 

\vspace{0.4 cm}
\hrule
\vspace{0.4 cm}

\normalsize

\section{Introduction}
Lyophilization, also known as freeze drying, is a process used to increase the stability of biotherapeutics in pharmaceutical manufacturing \cite{Liapis_1994_Original}. In recent studies, lyophilization has been shown to provide long-term stability for mRNA vaccines, allowing these vaccines to be stored at higher temperature while preserving their functionality \cite{Muramatsu_2022_mRNA,Meulewaeter_2023_mRNA}. This promising advancement could play an important role in future mRNA-based therapeutic manufacturing, in particular vaccine distribution in regions where a cold supply chain is lacking.

Three stages of lyophilization comprise (1) freezing, (2) primary drying, and (3) secondary drying, respectively. In freezing, the product and liquid solvent (usually water) are frozen, in which the free water becomes ice crystals, whereas the bound water retains its liquid state and is adsorbed to the organic material between the ice crystals \cite{Fissore_2018_Review}. In primary drying, the free water (in the form of ice crystals) is removed via sublimation \cite{Pisano_2010_control}. Subsequently, secondary drying is conducted at higher temperature to remove the bound water via desorption \cite{Sadikoglu_1997_Modeling}. The stability of a lyophilized product is significantly influenced by the amount of bound water, and so monitoring the concentration of bound water is highly important \cite{Fissore_2018_Review}. One of the most common techniques is the Karl Fischer titration, which requires sampling of the vial for offline measurements \cite{Pikal_2005_Model,Fissore_2018_Review}. To avoid process interruption, some online or non-invasive techniques such as near-infrared (NIR) spectroscopy \cite{Ikeda_2022_NIRmonitor} and tunable diode laser absorption spectroscopy (TDLAS) \cite{Schneid_2011_TLDASmonitor} have been proposed. Detailed discussion of tools for the monitoring of secondary drying can be found in Ref.\ \cite{Fissore_2018_Review}.

Instead of direct measurement, a state observer (aka state estimator, observer, estimator) can be used to estimate states that are not measured \cite{Luenberger1971Observer}; the process is known as state estimation. A well-designed observer can replace expensive and complicated sensors in the system, reducing the total cost and complexity of operation. The principle of state estimation is to combine available measurement data of some states with the physics of a system represented by a mechanistic model, and use that information to estimate the unmeasured states. A variety of mechanistic models for lyophilization are available \cite{Litchfield_1979_Model,Liapis_1994_Original,Mascarenhas_1997_FEMmodel,Sadikoglu_1997_Modeling,Sheehan_1998_Modeling,Veraldi_2008_Parameters,Nastaj_2009_MFD,Fissore_2011_SecDryingMonitor,Fissore_2015_Review}, which establishes a solid foundation for constructing a reliable observer.

Various state observers have been proposed and successfully implemented in chemical processes \cite{Ali_2015_ReviewOfChemicalProcessObserver}. In the context of lyophilization, state observers have been extensively studied and applied to the primary drying step, which aims at estimating the temperature, interface position (amount of ice), and relevant parameters such as the heat transfer coefficient \cite{Velardi_2009_ObsPrimary,Velardi_2010_ObsPrimary,Bosca_2011_ObsPrimary,Bosca_2013_ObsPrimary,Dragoi_2013_ANNobs,Bosca_2015_ObsPrimary,Bosca_2017_ObsPrimary,Fissore_2017_ObsPrimary}. Besides monitoring, some other applications such as process  optimization \cite{Bosca_2016_ObsOpt} and control \cite{Fissore_2008_ObsControl,Barresi_2009_ObsControl} have been demonstrated. However, applications of state estimation to secondary drying are very limited. The only literature that proposed a state estimation-like strategy for secondary drying is Ref.\ \cite{Fissore_2011_SecDryingMonitor}; the technique is referred to as a soft sensor that requires measurement of the desorption flux for estimating the residual moisture. The procedure described in the aforementioned work does not exploit the mathematical structure of a state observer; the key idea is to iteratively solve the optimization to find the moisture content that matches the measured desorption rate. This technique requires a set of equipment for measuring the desorption flux, which is generally available in pilot-scale lyophilizers.

In this article, a new technique is proposed for real-time estimation of bound water concentration during desorption, and is applied to the secondary drying step in lyophilization. The technique relies on a state observer that estimates the concentration of bound water by using the temperature measurement and mechanistic understanding of heat transfer and desorption kinetics. The proposed observer is extensively tested with various simulations and experiments. Since accurate bound water measurement is not trivial and usually involves complex equipment and procedures \cite{Joardder_2019_BWmeasurement}, our observer is formulated such that the only input required is temperature measurement, which is straightforward and very common in every step of lyophilization \cite{Fissore_2018_Review,Harguindeguy_2022_IR}, allowing for the simplest setup and operation compared to any other methods. The proposed framework can also be easily and systematically extended to other desorption-related processes.

This article is organized as follows. Section \ref{sec:Model} describes the lyophilization system and derives the corresponding mechanistic model. Section \ref{sec:Observer} discusses the design and implementation of our state observers. Section \ref{sec:Results} presents several simulation and experimental results to demonstrate the performance of the observers. Finally, Section \ref{sec:Conclusion} summarizes the study.

\section{Mechanistic Modeling} \label{sec:Model}
\begin{figure} [ht!]
\centering
    \includegraphics[scale=1.1]{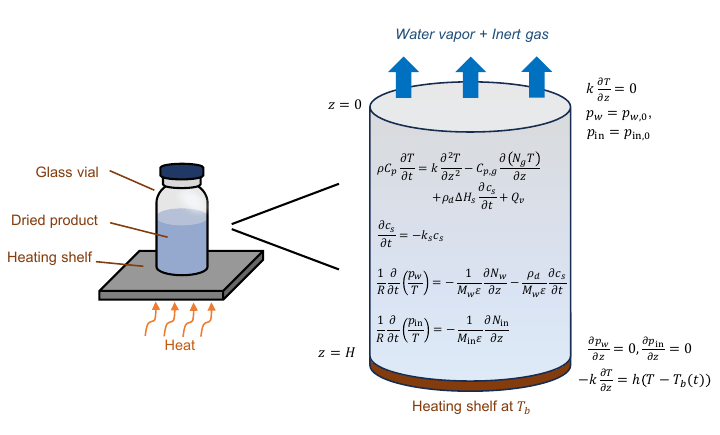}
    \caption {Schematic diagram showing a mechanistic model for the secondary drying step in lyophilization.} 
    \label{fig:Schematic}      
\end{figure}

A mechanistic model is an important element in a state observer as it contains the knowledge about the physics of a system. This work considers lyophilization of unit does, in which the product is introduced into a number of vials prior to being lyophilized. Our model is formulated in the rectangular coordinate system by considering one spatial dimension ($z$) and time ($t$) as pictorially shown in Fig.\ \ref{fig:Schematic}. For the drying steps in lyophilization, 1D modeling is nearly always used because its accuracy is comparable to that of multidimensional modeling while being much computationally cheaper and less complicated, whereas 0D modeling (lumped capacity) is sometimes not sufficiently accurate \cite{Yoon_2021_0D1D3DModeling}.

In secondary drying, there are three important transport phenomena, namely (1) heat transfer within the dried product, (2) desorption of the water vapor from the surface of the dried product, and (3) mass transfer of the water vapor in the pores of the dried product \cite{Litchfield_1979_Model,Liapis_1994_Original,Mascarenhas_1997_FEMmodel,Sadikoglu_1997_Modeling,Sheehan_1998_Modeling,Veraldi_2008_Parameters}. 

The first part of the model describes the heat transfer. The energy balance of the dried product is given by \cite{Litchfield_1979_Model,Liapis_1994_Original,Mascarenhas_1997_FEMmodel,Sadikoglu_1997_Modeling,Sheehan_1998_Modeling,Nastaj_2009_MFD} 
\begin{equation} \label{eq:energybalance}
    \rho C_p\dfrac{\partial T}{\partial t} = k\dfrac{\partial^2 T}{\partial z^2} - C_{p,g} \dfrac{\partial \left(N_gT\right)}{\partial z}  + \rho_d\Delta H_s\dfrac{\partial c_s}{\partial t} + Q_v, \qquad t>0,
\end{equation}
where $T(z,t)$ is the product temperature, $c_s(z,t)$ is the bound water concentration (aka moisture content, residual water), $\Delta H_s$ is the enthalpy of desorption, $\rho$ is the effective density, $\rho_d$ is the density of the dried product, $k$ is the effective thermal conductivity, $C_p$ is the effective heat capacity, $H$ is the height of the dried product, $N_g$ is the total mass flux of the gas (water vapor and inert gas), and $C_{p,g}$ is the heat capacity of the gas. Here effective parameters consider the properties of both solid and gas in the pores, the subscript $d$ denotes parameters for the dried product only (solid and vacuum), the subscript $g$ denotes parameters for the gas phase only, and the subscript $s$ denotes parameters related to desorption. 

The additional term $Q_v$ in \eqref{eq:energybalance} describes the effect of microwave irradiation, which provides volumetric heating to the product \cite{Fayaz_2015_VolumetricHeating,Abdelraheem_2022_Uniformity}. This study focuses on conventional lyophilization, i.e., no microwave, and so this term is set to 0 by default. Nevertheless, it is worth noting that our model and observer are designed to accommodate microwave lyophilization as well (see Section \ref{sec:Microwave}). The physics of microwave heating is related to dielectric heating, which is influenced by the electric field strength, the microwave frequency, and the dielectric loss factor of each material; detailed discussion can be found in Refs. \cite{Nastaj_2009_MFD,Wang_2020_MircowaveModel,Abdelraheem_2022_Uniformity}. Nevertheless, this information is not always readily available for all systems, so we model the microwave heating as a single composite term $Q_v$ in this case.

The bottom surface of the dried product is heated by the heating shelf, following Newton's law of cooling, 
\begin{equation}\label{eq:boundary_bottom}
    -k\dfrac{\partial T}{\partial z}(H,t) = h\left(T(H,t)-T_b(t)\right), \qquad t>0,
\end{equation}
where $h$ is the heat transfer coefficient at the bottom and $T_b(t)$ is the bottom shelf temperature. Typically, the shelf temperature increases linearly as a function of time, 
\begin{equation}\label{eq:shelf_temp}
    T_{b}(t) = rt + T_{b,0},
\end{equation}
where $T_{b,0}$ is the initial shelf temperature and $r$ is the temperature ramp-up rate. After reaching the maximum temperature $T_{b,\textrm{max}}$, the shelf temperature is kept constant at that value.\footnote{The methods apply for general $T_b(t)$.} At the top surface, heat transfer is negligible compared to the bottom surface, and thus the boundary condition is
\begin{equation} \label{eq:boundary_top}
    k\dfrac{\partial T}{\partial z} (0,t) = 0, \qquad  t>0.
\end{equation}
The initial temperature of the dried region is assumed to be spatially uniform at $T_0$,
\begin{equation} \label{eq:initial_temp}
    T(z,0) = T_0,  \qquad  0\leq z \leq H.
\end{equation}
As secondary drying takes place after primary drying, $T_0$ can be set to the sublimation temperature. 

The second part of the model concerns the desorption of bound water. It has been widely accepted in the literature that the linear driving force model can accurately predict the dynamics of bound water desorption despite being one of the simplest adsorption/desorption models \cite{Litchfield_1979_Model,Liapis_1994_Original,Sadikoglu_1997_Modeling,Sheehan_1998_Modeling,Sircar_2000_LDF,Veraldi_2008_Parameters,Fissore_2011_SecDryingMonitor,Fissore_2015_Review}. Hence, the desorption kinetics of bound water is described by 
\begin{equation} \label{eq:desorption_LDF}
     \dfrac{\partial c_\textrm{s}}{\partial t} = k_s(c^*_\textrm{s}-c_\textrm{s}),
\end{equation}
where $c^*_\textrm{s}$ is the equilibrium concentration of bound water and $k_s$ is the rate constant for desorption that exhibits Arrhenius temperature dependence \cite{Liapis_1994_Original,Sadikoglu_1997_Modeling,Fissore_2015_Review}
\begin{equation} \label{eq:rate_constant}
     k_s = Ae^{-E_a/RT},
\end{equation}
where $A$ is the frequency factor (aka collision frequency), $E_a$ is the activation energy, and $R$ is the gas constant. It is very common in the literature to set $c^*_\textrm{s}=0$. This simplification produces practically small errors as shown in Ref.\ \cite{Sadikoglu_1997_Modeling} and avoids the need for equilibrium data and detailed knowledge about the solid structure \cite{Fissore_2015_Review}. Thus, the final equation is 
\begin{equation} \label{eq:desorption}
     \dfrac{\partial c_s}{\partial t} = -k_sc_s.
\end{equation}
The initial concentration of bound water in secondary drying is assumed to be uniform at $c_{s,0}$,
\begin{equation} \label{eq:initial_conc}
    c_s(z,0) = c_{s,0},  \qquad  0\leq z \leq H.
\end{equation}
Mathematically, the main difference between the original equation \eqref{eq:desorption_LDF} and simplified version \eqref{eq:desorption} is that the final concentration approaches the equilibrium value for the former and approaches 0 for the latter. This work relies on the simplified equation \eqref{eq:desorption} due to its practicality and adequate accuracy, but the proposed state observer and relevant framework can be applied to both cases because the overall mathematical structure does not change. The original equation  \eqref{eq:desorption_LDF} and equilibrium data might be needed when there is evidence showing that the equilibrium concentration is significantly high or strict control at the very low concentration region is required.

The last part of the model focuses on the mass transfer of gas/vapor in the pores of the dried product, which usually consists of water vapor $(w)$ and inert gas (in). The continuity equations, assuming ideal gas behaviors for both components, are \cite{Litchfield_1979_Model,Liapis_1994_Original,Mascarenhas_1997_FEMmodel,Sadikoglu_1997_Modeling,Sheehan_1998_Modeling,Veraldi_2008_Parameters,Nastaj_2009_MFD}
\begin{gather}
\frac{1}{R}\frac{\partial}{\partial t} \!\left(\frac{p_w}{T}\right) = -\frac{1}{M_w\varepsilon} \frac{\partial N_w}{\partial z} - \frac{\rho_d}{M_w \varepsilon}\frac{\partial c_s}{\partial t}, \label{eq:water_massbalance} \\  
\frac{1}{R}\frac{\partial}{\partial t} \!\left(\frac{p_\textrm{in}}{T}\right) = -\frac{1}{M_\textrm{in}\varepsilon} \frac{\partial N_\textrm{in}}{\partial z},  \label{eq:inert_massbalance}
\end{gather}
where $p(z,t)$ is the partial pressure, $N(z,t)$ is the mass flux (mass flow rate per cross sectional area), $M$ is the molar mass, $\varepsilon$ is the porosity, and the subscripts $w$ and `in' denote the water vapor and inert gas, respectively. The expression for $N$ is usually modeled by the dusty-gas model \cite{Litchfield_1979_Model,Mascarenhas_1997_FEMmodel,Sadikoglu_1997_Modeling,Sheehan_1998_Modeling,Veraldi_2008_Parameters}. Note that the total mass flux $N_g$ in \eqref{eq:energybalance} is $N_w + N_\textrm{in}$. The initial conditions for both components are
\begin{gather}
p_w(z,0) = p_{w,0},  \qquad  0\leq z \leq H, \label{eq:initial_water} \\  
p_\textrm{in}(z,0) = p_{\textrm{in},0},  \qquad  0\leq z \leq H, \label{eq:initial_inert}   
\end{gather}
where $p_{w,0}$ and $p_\textrm{in,0}$ are usually defined by the condenser located downstream of the lyophilizer. The boundary conditions are
\begin{gather}
p_w(0,t) = p_{w,0},  \qquad t>0, \label{eq:top_water} \\  
p_\textrm{in}(0,t) = p_{\textrm{in},0},  \qquad t>0, \label{eq:top_inert}   \\
\frac{\partial p_w}{\partial z}(H,t) = 0 ,  \qquad t>0, \label{eq:bottom_water} \\  
\frac{\partial p_\textrm{in}}{\partial z}(H,t) = 0 ,  \qquad t>0. \label{eq:bottom_inert}  
\end{gather}
 
The main objective of secondary drying is to remove bound water; hence, the concentration of bound water $c_s$ is usually the variable of interest \cite{Sadikoglu_1997_Modeling,Fissore_2011_SecDryingMonitor,Fissore_2015_Review,Fissore_2018_Review}. The concentration of bound water can be predicted accurately with the energy balance and desorption kinetics equations as described and shown in Refs.\ \cite{Fissore_2015_Review,Sahni_2017_Simplified,Yoon_2021_0D1D3DModeling}, and so recent models usually omit the mass transfer of gas/vapor in the pores of the dried product, i.e., the last part of the model \eqref{eq:water_massbalance}--\eqref{eq:bottom_inert}. This approach simplifies the model equations, parameter estimation, and observer/control design, without significant loss in accuracy.

In this work, the above mechanistic model is simulated numerically. The model equations are spatially discretized using the finite volume method as explained in Appendix \ref{app:Model_FVM}. The final discretized equations can be written as 
\begin{equation} \label{eq:FinalODEs}
    \dfrac{d\mathbf{x}}{dt} = \mathbf{F(x)} + \mathbf{Bu},
\end{equation}
with the state $\mathbf{x}$ and manipulated variable $\mathbf{u}$ defined as
\begin{gather}
    \mathbf{x} = \!\begin{bmatrix}
     \mathbf{T}  \\
     \mathbf{c_s}
    \end{bmatrix}\!, \label{eq:state}\\
    \mathbf{u} =\! \begin{bmatrix}
         T_b  \\
         Q_v
    \end{bmatrix}\!,
\end{gather}
where $\mathbf{T} \in \mathbb R^m$ collects the product temperatures ($T_1,\dots,T_m$), $\mathbf{c_s} \in \mathbb R^m$ collects the bound water concentrations $(c_{s,1},\dots,c_{s,m})$, $m$ is the number of grid points in the spatial domain, $\mathbf F \in \mathbb R^{2m}$ is the nonlinear vector function, and $\mathbf B \in \mathbb R^{2m\times2}$ is the corresponding matrix. To facilitate the observer design, \eqref{eq:FinalODEs} can be rewritten as
\begin{gather} 
    \dfrac{d \mathbf{T}}{dt} = \mathbf{F_T(\mathbf{T}, \mathbf{c_s})} + \mathbf{B_Tu}, \label{eq:FinalODETemp}\\
    \dfrac{d \mathbf{c_s}}{dt} =  \mathbf{F_c(\mathbf{T}, \mathbf{c_s}) + \mathbf{B_cu}}, \label{eq:FinalODEConc}
\end{gather}
where $\mathbf{F_T} \in \mathbb R^{m}, \mathbf{B_T} \in \mathbb R^{m\times2} $ represent the dynamics of the temperature part and $\mathbf{F_c} \in \mathbb R^{m}, \mathbf{B_c} \in \mathbb R^{m\times2}$ represent the dynamics of the concentration part. The finite volume method transforms the original partial differential equations (PDEs) into a system of ordinary differential equations (ODEs). The final ODEs \eqref{eq:FinalODETemp} and \eqref{eq:FinalODEConc} can be integrated by typical ODE solvers, in which MATLAB's \texttt{ode15s} is used in this work. This technique is known as the method of lines \cite{Schiesser_1991_MOL}. Lastly, define the average temperature and average concentration,
\begin{gather}
    T_\textrm{avg} = \frac{1}{m} \sum_{i=1}^{m} T_i, \label{eq:temp_avg} \\
    c_{s,\textrm{avg}} = \frac{1}{m} \sum_{i=1}^{m} c_{s,i}. \label{eq:conc_avg}    
\end{gather}

\section{State Observer} \label{sec:Observer}
A state observer (aka state estimator, observer, estimator) is a tool in control theory that is used for reconstructing the unmeasured states given the available measurements and mechanistic understanding of a system; the process is referred to as state estimation. Those unmeasured states could be internal states that cannot be measured or states that are difficult to measure. Measuring the temperature during lyophilization is relatively simple and accurate, so we design a state observer that uses the temperature measurement to estimate the concentration of bound water, the most important process variable in secondary drying.

State estimation is critical for process monitoring and control, in which the information of the unmeasured states is needed. Various state observers have been proposed and employed \cite{Ali_2015_ReviewOfChemicalProcessObserver}. The Luenberger observer \cite{Luenberger1971Observer} has a simple mathematical structure and is computationally efficient for both linear and nonlinear processes, which has resulted in its widespread use in various applications \cite{Srisuma_2023_1DCellThawing}.

\subsection{Mathematical structure of a state observer} \label{sec:ObsStructure}
Applying the Luenberger observer to the final model equation \eqref{eq:FinalODEs} results in
\begin{equation} \label{eq:Observer_1}
    \dfrac{d\hat{\mathbf{x}}}{dt} = \mathbf{F(\hat{x})} + \mathbf{Bu} + \mathbf{L}(\hat{\mathbf{y}} - \mathbf{y}),
\end{equation}
where $\hat{\mathbf{x}}\in \mathbb R^{2m}$ is the estimated state predicted by the observer, $\mathbf{y}$ is the measured outputs, $\hat{\mathbf{y}}$ is the estimated outputs, and $\mathbf{L}$ is the observer gain. Similar to the actual state defined by \eqref{eq:state}, the estimated state $\mathbf{x}$ is
\begin{equation}
    \hat{\mathbf{x}} = \begin{bmatrix}
     \hat{\mathbf{T}}  \\
     \mathbf{\hat{c}_s}
    \end{bmatrix}.
\end{equation}
where $\hat{\mathbf{T}}\in \mathbb R^{m}$ is the estimated temperature and $\mathbf{\hat{c}_s}\in \mathbb R^{m}$ is the estimated concentration. The measured output is the temperature profile of the product, so
\begin{gather}
    \mathbf{y} = \mathbf{T} + \mathbf{n} \label{eq:measurement}, \\
    \hat{\mathbf{y}} = \hat{\mathbf{T}} \label{eq:measurement_estimated},    
\end{gather}
where $\mathbf{n}\in \mathbb R^{m}$ is the sensor noise. The most important part of the observer is the observer gain $\mathbf{L} \in \mathbb R^{2m \times m}$, which directly affects the performance of the observer. Depending on the knowledge of the system, different strategies can be used to design the observer gain. 

To simplify the observer design, we separate the observer gain matrix $\mathbf{L}$ into two parts corresponding to the temperature and concentration, that is, 
\begin{gather}
    \mathbf{L} = \begin{bmatrix}
     \mathbf{L_T}  \\
     \mathbf {L_c}
    \end{bmatrix},   
\end{gather}
where $\mathbf{L_T}\in \mathbb R^{m \times m}$ is the observer gain for the temperature part and $\mathbf {L_c}\in \mathbb R^{m \times m}$ is the observer gain for the concentration part. Consequently, \eqref{eq:Observer_1} can be rewritten as
\begin{gather}
    \dfrac{d\hat{\mathbf{T}}}{dt} = \mathbf{F_T\left(\hat{T},\hat{c}_s\right)} +  \mathbf{B_Tu} + \mathbf{L_T}\left(\hat{\mathbf{T}} - \mathbf{T}\right), \label{eq:Observer_T} \\
    \dfrac{d\mathbf{\hat{c}_s}}{dt} = \mathbf{F_c\left(\hat{T},\hat{c}_s\right)} + \mathbf{B_cu} + \mathbf{L_c}\left(\hat{\mathbf{T}} - \mathbf{T}\right). \label{eq:Observer_cs}      
\end{gather}
Here the first part of the observer \eqref{eq:Observer_T} estimates the product temperature, while the second part \eqref{eq:Observer_cs} estimates the residual moisture. Separating the observer gains allows each part of the observer to be designed separately while still respecting the coupling of the states in the original model.

The final step is to design the observer gains $\mathbf{L_T}$ and $\mathbf{L_c}$. The current observer gains $\mathbf{L_T}$ and $\mathbf{L_c}$ are $m$$\times$$m$ matrices, which leaves many degrees of freedom in the observer design. Therefore, we parameterize the observer gain matrices by 
\begin{gather}
    \mathbf{L_T} = L_T \mathbf{J}_m \label{eq:gain_T}, \\
    \mathbf{L_c} = L_c \mathbf{J}_m \label{eq:gain_cs},  
\end{gather}
where $L_T, L_c$ are the real scalars and $\mathbf{J}_m$ is an $m$$\times$$m$ matrix of ones. 
This parameterization suggests that the temperature measurement at each location contributes equally to the observer, leaving only two degrees of freedom for the design: $L_T$ and $L_c$. A well-designed observer should converge the estimated states to the true states fast compared to the time scale of the process. The convergence can be evaluated via the estimation errors defined as
\begin{gather}
    \mathbf{e_T} = \left|\hat{\mathbf{T}}-\mathbf{T}\right|, \\
    \mathbf{e_c} = \left|\mathbf{\hat{c}_s} - \mathbf{c_s}\right|,
\end{gather}
where $\mathbf{e_T}$ is the estimation error for temperature and $\mathbf{e_c}$ is the estimation error for concentration. A zero estimation error indicates the convergence of the estimated state.

\subsection{Modified state observer} \label{sec:ModObserver}
The state observer proposed in Section \ref{sec:ObsStructure} receives spatially distributed temperature measurement ${\mathbf{T}}$ and provides estimates of both temperature $\mathbf{\hat{T}}$ and concentration $\mathbf{\hat{c}_s}$ in real time. Although spatially distributed temperature measurement can be obtained using thermal imaging sensors, traditional lyophilization systems do not have such sensors. For example, a thermocouple used for temperature measurement is usually in contact with the bottom of the product, and so only the bottom temperature is available \cite{Veraldi_2008_Parameters,Fissore_2018_Review}. Therefore, we propose an alternative state observer for this scenario.

For convenience, we denote this alternative as a {\it modified state observer}. Modifying the original observer to take the bottom temperature measurement instead of the spatial temperature measurement results in the output vectors $\mathbf{y}$ and $\hat{\mathbf{y}}$
\begin{gather}
    y =  T_p + n \label{eq:measurement2}, \\
    \hat{y} = \hat{T}_p \label{eq:measurement2_estimated},    
\end{gather}
where the measurement noise $n$ is a real scalar and $T_p$ is the bottom temperature. In the state vector, $T_p$ corresponds to the last element of ${\mathbf{T}}$. As a result, the equations for this observer are
\begin{gather}
    \dfrac{d\hat{\mathbf{T}}}{dt} =  \mathbf{F_T\left(\hat{T},\hat{c}_s\right)} +  \mathbf{B_Tu} + \mathbf{L_T}\left(\hat{T}_p - T_p\right), \label{eq:Observer2_T} \\
    \dfrac{d\mathbf{\hat{c}_s}}{dt} = \mathbf{F_c\left(\hat{T},\hat{c}_s\right)} +  \mathbf{B_cu} + \mathbf{L_c}\left(\hat{T}_p - T_p\right), \label{eq:Observer2_cs}  
\end{gather}
where $\mathbf{L_T} \in \mathbb{R}^{m}$ and $\mathbf{L_c} \in \mathbb{R}^{m}$ are the observer gains. 
Similarly, we can parameterize $\mathbf{L_T}$ and $\mathbf{L_c}$ can be parameterized using the vector of ones (instead of the matrix of ones as used for the original state observer) such that the only design parameters are the real scalars $L_T$ and $L_c$.\footnote{Since the observer structure and equations are different, the values of $L_T$ and $L_c$ used for the state observer and modified observers are also different.}

Instead of bottom temperature measurement, some systems may have sensors installed at the top, and hence the temperature at the top surface is measured \cite{Colucci_2020_IR,Srisuma_2023_1DCellThawing}. In such cases, a similar procedure introduced in this section can be applied.

Most of the results in this article are based on the original state observer as it uses the most complete measurement information, with some results and discussion on the modified observer in Section \ref{sec:RealSystems}.

\subsection{Observer design strategies} \label{sec:DesignStrategies}
\begin{figure} [ht!]
\centering
    \includegraphics[scale=1]{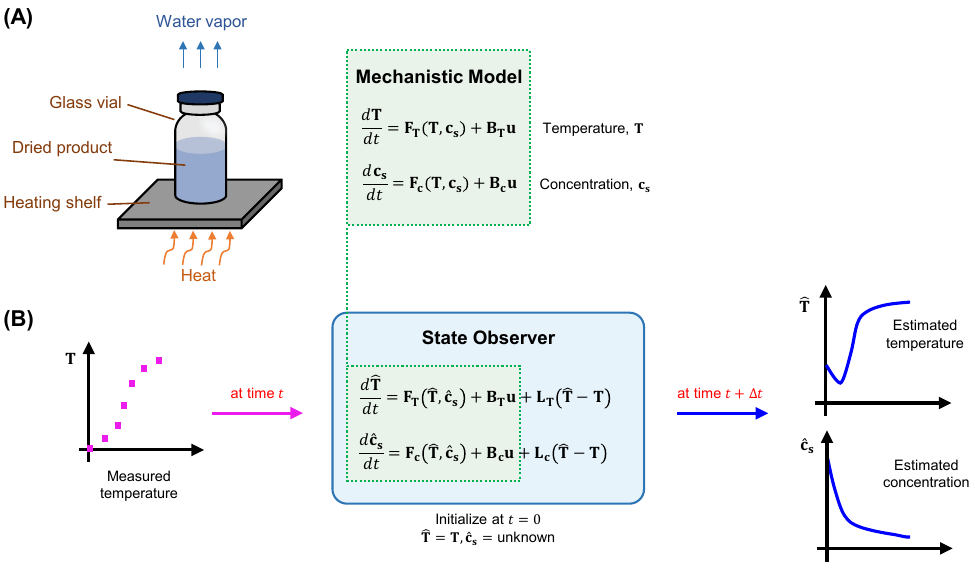}
   
   \vspace{-.3cm} 
   
   \caption {(A) Schematic diagram of the typical lyophilization process. In secondary drying, a glass vial that contains the dried product is heated such that the bound water is desorbed and removed from the product at the top. The mechanistic model is used to describe the heat transfer and desorption dynamics during secondary drying.(B) Structure of the proposed state observer. The observer receives the measured temperature and the control input (i.e., the shelf temperature and microwave power) at time step $t$, and estimates the temperature and concentration at the next time step. To initialize the observer at $t=0$, the estimated temperature can be set to the measured value. The initial concentration is completely unknown due to no measurement, so it can be set to some realistic values obtained from the literature or previous experiment.} 
    \label{fig:Observer}      
\end{figure}

The structure of the state observer for estimating the concentration of bound water is illustrated in Fig.\ \ref{fig:Observer}. The proposed observer has the physics of heat transfer and bound water desorption embedded in $\mathbf{F_T}$ and $\mathbf{F_c}$, which is given by the mechanistic model. The temperature measurement is fed to the observer terms represented by the observer gains $\mathbf{L_T}$ and $\mathbf{L_c}$. The observer combines the information from the mechanistic model and temperature measurement to converge the estimated concentration to the true values without the need for concentration measurement. 

The most important consideration for observer design is to ensure that the estimated states converge to the true states fast compared to the time scale of the process, but have slow enough dynamics that the state estimates are insensitive to measurement noise. The time scales of the estimated states are specified by the observer gains $\mathbf{L_T}$ and $\mathbf{L_c}$. Instead of attempting to search over all elements of the full observer gain matrices $\mathbf{L_T}$ and $\mathbf{L_c}$, we show in Sections \ref{sec:ObsStructure} and \ref{sec:ModObserver} that $\mathbf{L_T}$ and $\mathbf{L_c}$ be parameterized and rewritten as a product of the real scalar $L_T$ or $L_c$ and the matrix of ones, so the only design parameters are the real scalars $L_T$ and $L_c$. 

The design procedure for the observer gains $L_T$ and $L_c$ consists of two main steps. The first step relies entirely on the mechanistic model and simulation, where the model prediction represents the true state of the system. In this step, a series of simulations are run for different values of the observer gains and parameters to investigate the dynamics of the system and observer under various conditions, and that information is used to select the observer gains based on the overall performance of the observer. The second step takes the real data/measurement from experiment into account, and so the observer gains are fine-tuned to be specific to the real system. These two steps are denoted as {\it simulation-based observer design} and {\it experiment-based observer design}, respectively. The former allows for different operating conditions and noise profiles to be tested so that the resulting design can cover many possible scenarios, whereas the latter is primarily specific to the system where the real data are available. This practical two-step procedure allows the observer to be designed efficiently without any complicated mathematical analysis, and is widely used in industrial applications. For the system, $L_T$ and $L_c$ have the units of s$^{-1}$ and kg water/(kg solid\,$\cdot$\,K\,$\cdot$\,s), but only the magnitudes of $L_T$ and $L_c$ are reported to keep the plots simple and easy to visualize.

An alternative to the first step in observer design is to set the observer gains based on an analytical expression for the convergence time for the observer dynamics derived from the mathematical structure of the observer. To begin this analysis, first define the Jacobian of the nonlinear function $\mathbf{F}$ in \eqref{eq:FinalODEs} as
\begin{equation} \label{eq:Jacobian}
    \mathbf{F'} = \frac{\partial \mathbf{F}}{\partial \mathbf{x}},
\end{equation}
where $\mathbf{F'}$ can be calculated analytically or numerically. One of the simplest but efficient ways of analyzing a nonlinear state observer is to approximate the nonlinearities by linear equations, i.e., apply linearization. Linearization of the model equation \eqref{eq:FinalODEs} results in 
\begin{equation} \label{eq:FinalODEs_lin}
    \dfrac{d\mathbf{x}}{dt} = \mathbf{F_{ref}} + \mathbf{F'_{ref}(x-x_{ref})}  + \mathbf{Bu},
\end{equation}
where $\mathbf{F_{ref}} = \mathbf{F(x_{ref})}$ is the function $\mathbf{F}$ evaluated at $\mathbf{x_{ref}}$, $\mathbf{F'_{ref}} = \mathbf{F'(x_{ref})}$ is the Jacobian $\mathbf{F'}$ evaluated at $\mathbf{x_{ref}}$, and $\mathbf{x_{ref}}$ is the reference state. The average values of temperature and concentration between the initial and final times are used as the reference state, which is constant and uniform.\footnote{Although there is no real-time concentration measurement available, the average concentration used in this analysis can be obtained via offline measurement during the design and development phase. Alternatively, a literature value could be used.} With \eqref{eq:FinalODEs_lin}, the linearized state observer is
\begin{equation} \label{eq:Observer_1_lin}
    \dfrac{d\hat{\mathbf{x}}}{dt} =  \mathbf{F_{ref}} +\mathbf{F'_{ref}(\hat{x}-x_{ref})} + \mathbf{L}(\hat{\mathbf{y}} - \mathbf{y}) + \mathbf{Bu}.
\end{equation}
Subtracting \eqref{eq:FinalODEs_lin} from \eqref{eq:Observer_1_lin} yields
\begin{equation} \label{eq:error_lin}
    \dfrac{d}{dt} (\hat{\mathbf{x}} - \mathbf{x}) =  \mathbf{F'_{ref}}(\hat{\mathbf{x}} - \mathbf{x}) + \mathbf{L}(\hat{\mathbf{y}} - \mathbf{y}).
\end{equation}
In this case, it is useful to write $\mathbf{y}$ and $\hat{\mathbf{y}}$ as a function of $\mathbf{x}$, that is,
\begin{gather}
    \mathbf{y} = \mathbf{Cx} + \mathbf{n} \label{eq:measurement_new}, \\
    \hat{\mathbf{y}} = \mathbf{C}\hat{\mathbf{x}}  \label{eq:measurement_estimated_new},    
\end{gather}
where 
\begin{equation}
    \mathbf{C} = \begin{bmatrix}
        \mathbf{I}_m & \mathbf{0}_{m,m}
    \end{bmatrix},
\end{equation}
$\mathbf{I}_m$ is an $m$$\times$$m$ identity matrix, and $\mathbf{0}_{m,m}$ is an $m$$\times$$m$ zero matrix. With this definition, it is easy to see that \eqref{eq:measurement_new} and \eqref{eq:measurement_estimated_new} are identical to \eqref{eq:measurement} and \eqref{eq:measurement_estimated}. Substituting  \eqref{eq:measurement_new} and \eqref{eq:measurement_estimated_new} into \eqref{eq:Observer_1_lin} followed by rearranging gives that
\begin{equation} \label{eq:error_lin2}
    \dfrac{d}{dt} \left(\hat{\mathbf{x}} - \mathbf{x}\right) =  \left(\mathbf{F'_{ref}}+\mathbf{LC}\right)\left(\hat{\mathbf{x}} - \mathbf{x}\right) - \mathbf{Ln},
\end{equation}
which is a linear ODE that gives the criterion for observer design. The eigenvalues of the matrix $\mathbf{F'_{ref}}+\mathbf{LC}$ characterize the dynamics of the observer. For the estimation error $\hat{\mathbf{x}} - \mathbf{x}$ to converge to zero, the observer gains $L_T$ and $L_c$ must be chosen such that the real parts of all eigenvalues of the matrix $\mathbf{F'_{ref}}+\mathbf{LC}$ are negative. Besides, the estimation error should decay significantly faster than the slowest time scale of the process. Oscillation should also be minimized, which is governed by the imaginary parts of all eigenvalues. 

The values of the $L_T$ and $L_c$ in \eqref{eq:error_lin2} can be systematically selected to give a desired convergence time for the observer dynamics, which can be computed from its slowest time scale $\tau$. Denote the eigenvalues of the matrix $\mathbf{F'_{ref}}+\mathbf{LC}$ from the fastest to the  slowest as $\lambda_1, \lambda_2, \dots, \lambda_{2m}$, in which $\lambda_1$ is the fastest and $\lambda_{2m}$ is the slowest. As $\mathbf{F'_{ref}}+\mathbf{LC}$ is a $2m$$\times$$2m$ matrix, the matrix has $2m$ eigenvalues in total. The slowest time constant can be estimated from the slowest eigenvalue:
\begin{equation}
    \tau = \frac{1}{\left|\mathrm{Re}(\lambda_{2m})\right|}.
\end{equation}
In this particular application, a much more accurate approximation is
\begin{equation} \label{eq:timeconstant}
    \tau = \frac{1}{\left|\mathrm{Re}(\lambda_{m+1})\right|},
\end{equation}
which is justified in Appendix \ref{app:TimeConstant} by a detailed mathematical analysis on the effects of all eigenvalues. The convergence time is defined as the time required for the estimation error to be less than 2\% of the initial state error, which is 4$\tau$.\footnote{This approach to selecting time scales in observer design is similar to that commonly used in pole placement and internal model control applications \cite{Brasch1970,Morari1989}.} 
This approximation is sufficiently accurate for all parameter values for the mechanistic model, numerical methods, and state observer used in this work, that is, for which the observer dynamics are stable and not oscillating. This analysis is based on a linearized version of the observer, and so is an approximation. Nevertheless, this approximation is sufficiently good for this application, including for gain scheduling, as shown in Section \ref{sec:Noise}.

\section{Results and Discussion} \label{sec:Results}
In this work, the mechanistic model is used in various applications with different operating conditions, which requires a number of input parameters. The default parameter values are given in Table \ref{Tab:Parameters}. Parameter values different from those reported in Table \ref{Tab:Parameters} are stated explicitly in each specific section or case study. 

\begin{table}[ht!]
\caption{Default parameters for simulations.}
\renewcommand{\arraystretch}{1.2}
\label{Tab:Parameters}
\centering
\begin{tabular}{| c | c | c |c |}
\hline
\textbf{Parameter} & \textbf{Value} & \textbf{Unit} & \textbf{Reference} \\ 
\hline
$\rho$  & 215  & kg/m$^3$ &   \cite{Sadikoglu_1997_Modeling}  \\ 
$\rho_d$ & 212.21 & kg/m$^3$ &   \cite{Sadikoglu_1997_Modeling}  \\
$k$ & 0.217 & W/(m$\cdot$K)  &   \cite{Sadikoglu_1997_Modeling} \\  
$C_p$ & 2,590  & J/(kg$\cdot$K)  &   \cite{Sadikoglu_1997_Modeling}  \\ 
$C_{p,g}$ & 1,617  & J/(kg$\cdot$K)  &    \cite{Sadikoglu_1997_Modeling} \\ 
$\Delta H_s$ & $2.68$$\times$$10^6$  &  J/kg &   \cite{Sadikoglu_1997_Modeling}\\
$E_a$ & 8,316  &  J/mol &   \cite{Pikal_2005_Model}\\
$A$ & 3.34$\times$$10^{-3}$  &  s$^{-1}$ &   \cite{Pikal_2005_Model}\\
$h$ & 30 & W/(m$^2\cdot$K) &    \cite{Sheehan_1998_Modeling} \\ 
$T_0$ & 241.15 &  K &  \cite{Gitter_2018_Experiment}\\
$T_{b,0}$ & 253.15 &  K & \cite{Gitter_2018_Experiment} \\
$T_{b,\textrm{max}}$ & 313.15 & K &  \cite{Sadikoglu_1997_Modeling} \\
$c_{s,0}$ & 0.2059 &  kg water/kg solid & \cite{Pikal_2005_Model} \\
$r$ & 0.2 &  K/min &   \cite{Gitter_2019_Parameters} \\
$Q_v$ & 0 &  W/m$^3$ &   -- \\
$H$ & 2 & cm &   \cite{Sadikoglu_1997_Modeling} \\ 
$R$ & 8.314 & J/(mol$\cdot$K) &   -- \\ 
$m$ & 20 & -- &   -- \\ 
\hline
\end{tabular}
\renewcommand{\arraystretch}{1}
\end{table}

Despite a large number of parameters in the model, there are two sets of parameters that are critical and thus should be estimated from data, namely the (1) heat transfer coefficient $h$ and (2) parameters related to the desorption rate constant $A, E_a$. The heat transfer coefficient should be estimated from temperature data; the typical value of a heat transfer coefficient for lyophilization applications ranges from 1 to 100 W/(m$^2\cdot$K) \cite{Hottot_2005_HTC,Hottot_2006_Parameters}. The activation energy for water desorption $E_a$ varies significantly among different systems; the values were reported between 5$\times$10$^3$ J/mol up to 5$\times$10$^4$ J/mol \cite{Pikal_2005_Model,Li_2010_Ea,Fissore_2015_Review,Kosasih_2022_Ea}. The frequency factor $A$ varies by many orders of magnitude \cite{Sadikoglu_1997_Modeling,Pikal_2005_Model,Fissore_2015_Review}, and so there is no typical value suggested. Both the activation energy and frequency factor should be estimated from bound water concentration data. The aforementioned bounds and typical values are used for parameter estimation in this work to ensure physically reasonable parameter values.

\subsection{Model validation} \label{sec:ModelVal}
Since the mechanistic model is a basis for state observer and control design, this section extensively validates the proposed model with three different datasets from the literature to ensure that our model provides an accurate prediction of the bound water concentration and product temperature. Table \ref{Tab:ParametersVal} lists the parameter values specific to the three case studies used for model validation; other parameters are based on the default values given in Table \ref{Tab:Parameters}. 

\begin{table}[ht!]
\caption{Specific parameters for the model validation.}
\renewcommand{\arraystretch}{1.2}
\label{Tab:ParametersVal}
\centering
\begin{tabular}{| c | c | c |c | c |}
\hline
\textbf{Case}&\textbf{Parameter} & \textbf{Value} & \textbf{Unit} & \textbf{Reference/Note} \\ 
\hline 
\multirow{3}{*}{1} & $E_a$ & 5,000 &  J/mol &  Estimated from data \\
& $A$ &  7.1$\times$10$^{-4}$  &  s$^{-1}$ &   Estimated from data \\
 & $c_{s,0}$ & 0.6415 &  kg water/kg solid & \cite{Sadikoglu_1997_Modeling} \\
\hline
\multirow{3}{*}{2} & $E_a$ & 5,700 &  J/mol &  Estimated from data \\
& $A$ &  1$\times$10$^{-3}$  &  s$^{-1}$ &   Estimated from data \\
 & $c_{s,0}$ & 0.6415 &  kg water/kg solid & \cite{Sheehan_1998_Modeling} \\
 \hline
\multirow{7}{*}{3} & $E_a$ & 5,920 &  J/mol &  \cite{Fissore_2015_Review} \\
& $A$ & 1.2$\times$$10^{-3}$  &  s$^{-1}$ &   Estimated from data \\
& $h$ & 7 & W/(m$^2\cdot$K) &   Estimated from data  \\ 
& $T_0$ & 264.09 &  K &  \cite{Fissore_2015_Review} \\
& $T_{b,0}$ & 264.09 &  K & Assumed to be $T_0$  \\
& $T_{b,\textrm{max}}$ & 312 &  K &  Estimated from data  \\
& $c_{s,0}$ & 0.0603 &  kg water/kg solid & \cite{Fissore_2015_Review} \\
& $r$ & 0.5 &  K/min &  Estimated from data \\
& $H$ & 0.0102 &  m & Calculated from parameters in Ref.\ \cite{Fissore_2015_Review} \\ \hline
\end{tabular}
\renewcommand{\arraystretch}{1}
\end{table} 

\begin{figure} [ht!]
\centering
    \includegraphics[scale=1]{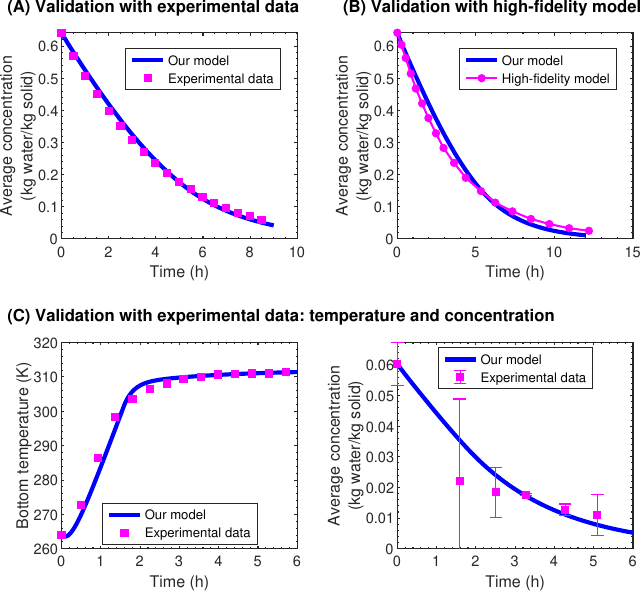}
    \caption {(A) Comparison between the concentration of bound water predicted by our model and the experimental data obtained from \cite{Sadikoglu_1997_Modeling}. The original data reported as the total mass of water are normalized by the total mass of solid. (B) Comparison between the concentration of bound water predicted by our model and the high-fidelity model by \cite{Sheehan_1998_Modeling}. The original data reported as the total mass of water are normalized by the total mass of solid. (C) Comparison between our model prediction and the experimental data from \cite{Fissore_2015_Review} for both temperature and concentration.} 
    \label{fig:Validation_all}      
\end{figure}

The first dataset, denoted Case 1, is obtained from the experimental data presented in Ref.\ \cite{Sadikoglu_1997_Modeling}, where the time profile of the total mass of residual water (bound water) during secondary drying was reported. Our mechanistic model can accurately predict the concentration of bound water, with the maximum error of about 0.01 kg water/kg solid (Fig.\ \ref{fig:Validation_all}A). The concentration decreases exponentially following the linear driving force model.

The second dataset (Case 2) is the simulation result obtained from the high-fidelity model proposed by Ref.\ \cite{Sheehan_1998_Modeling}; the model simulates simultaneous heat and mass transfer in two dimensions. Our model prediction agrees well the result obtained from the high-fidelity model (Fig.\ \ref{fig:Validation_all}B), indicating that our simplified model, which simulates the system in 1D and omits mass transfer equations, is sufficiently accurate to be used for predicting the evolution of residual water during secondary drying, agreeing with the observation by Ref.\ \cite{Yoon_2021_0D1D3DModeling}.

The final dataset (Case 3) is obtained from Ref.\ \cite{Fissore_2015_Review}, where the time profiles of both residual water and product temperature were reported. Our model can precisely simulate the product temperature (Fig.\ \ref{fig:Validation_all}c). The model can also reasonably predict the concentration of bound water despite high uncertainty in the reported measurement.  

The above three case studies show that our proposed model provides accurate prediction of the bound water concentration and product temperature during secondary drying, in comparison to experimental data and a model with higher fidelity.

\subsection{Simulation-based observer design} \label{sec:SimDesign}
This section demonstrates the simulation-based observer design, with the default model parameters in Table \ref{Tab:Parameters}. As introduced in Section \ref{sec:DesignStrategies}, a practical design technique is to run a series of simulations for different values of the observer gains. Alternatively, the observer gains can be systematically selected from mathematical analysis of the observer \eqref{eq:Jacobian}--\eqref{eq:timeconstant} to produce a desired convergence time for the estimation error.

\begin{figure} [ht!]
\centering
    \includegraphics[scale=1]{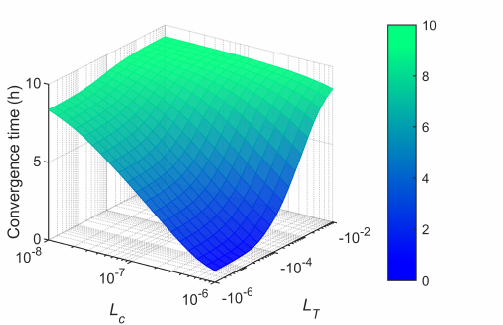}  
    \caption{Design space showing the times required for the estimated concentration to converge to the true value at different pairs of the observer gains $L_T$ and $L_c$. Values of $L_T$ and $L_c$ outside this design space could lead to severe oscillation or divergence.} 
    \label{fig:Obs0}      
\end{figure}

Figure \ref{fig:Obs0} shows the convergence times for orders of magnitude ranges in observer gains; the convergence time can be estimated using \eqref{eq:timeconstant} or directly calculated from simulation. Within this design space, the convergence time ranges from 0.7 to 9.3 h. For the process time scale of about 10 h, we select the observer gains $L_T = -1$$\times$$10^{-6}$ and $L_c = 5$$\times$$10^{-7}$ to give a convergence time of about 2 h (dark blue region of Fig.\ \ref{fig:Obs0}), i.e., 5 times faster than the process dynamics. 

\begin{figure} [ht!]
\centering
    \includegraphics[scale=1]{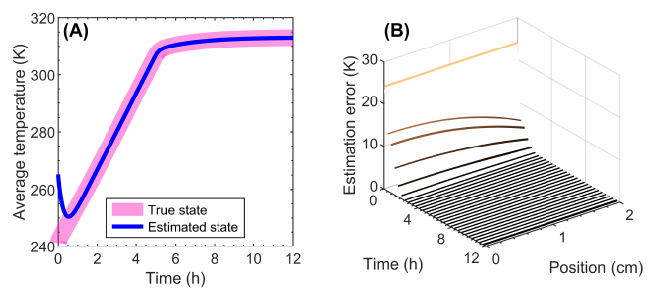}
    \caption{Convergence of the estimated product temperature for the selected observer gains $L_T = -1$$\times$$10^{-6}$ and $L_c = 5$$\times$$10^{-7}$. Panel A shows the evolution of the estimated average temperature compared to the true value. Panel B shows the spatiotemporal evolution of the estimation error. The initial estimated temperature is set to be higher than the actual value by about 10\% to demonstrate convergence.} 
    \label{fig:Obs1T}      
\end{figure}

\begin{figure} [ht!]
\centering
    \includegraphics[scale=1]{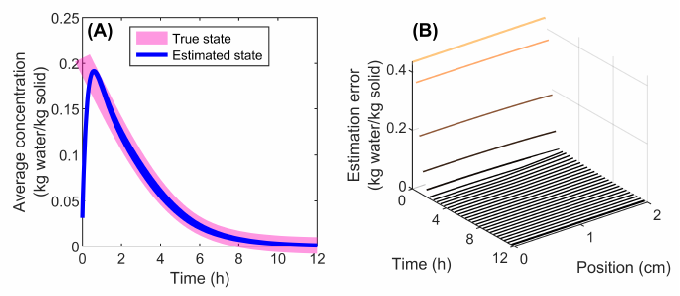}
    \caption{Convergence of the estimated concentration of bound water for the selected observer gains $L_T = -1$$\times$$10^{-6}$ and $L_c = 5$$\times$$10^{-7}$. Panel A shows the evolution of the estimated average concentration compared to the true value. Panel B shows the spatiotemporal evolution of the estimation error. The initial estimated concentration is set to 0.0314 kg water/kg solid, the minimum value reported in the literature.} 
    \label{fig:Obs1c}      
\end{figure}

With the selected observer gains, the estimated temperature quickly converges to the true value without any oscillation (Fig.\ \ref{fig:Obs1T}A), with the estimation error decreasing to 0 at around 2 h (Fig.\ \ref{fig:Obs1T}B).  A similar behavior can be observed for the concentration (Fig.\ \ref{fig:Obs1c}), indicating that the observer design is appropriate. Physically, when the estimated temperature is higher than the actual value, the observer term $\mathbf{L_T}(\hat{\mathbf{T}} - \mathbf{T})$ should provide negative feedback to the observer to reduce the temperature, so $L_T$ is negative. When the estimated temperature is higher than the actual value, it implies that the estimated desorption rate is too low. Therefore, the observer term $\mathbf{L_c}(\hat{\mathbf{T}} - \mathbf{T})$ should provide positive feedback to the observer to increase the estimated concentration, and so $L_c$ is positive.

Another important aspect of the observer is initialization. For the estimated temperature, it is logical to set the initial condition to be equal to the measured temperature. In Fig.\ \ref{fig:Obs1T}, the initial estimated temperature is set to be higher than the measured value by 10\% to demonstrate the convergence. The initial estimated concentration is, however, unknown due to no real-time concentration measurement. In Fig.\ \ref{fig:Obs1c}, the observer is initialized with the minimum concentration (0.0314 kg water/kg solid) reported in the literature \cite{Sadikoglu_1997_Modeling,Pisano_2012_ObsSecondary} to demonstrate the convergence under the worst-case scenario. In practice, the initial estimated concentration could be set to some more realistic value specific to that system or experiment. This information should be obtained during the design and development phase, which could result in faster convergence. In the later sections of this work, the initial estimated concentration is set to 0.0314 kg water/kg solid and the initial estimated temperature is set to the measured value unless otherwise specified so that the observer performance is analyzed on the same basis.

In the later sections of this work, $L_T = -1$$\times$$10^{-6}$ and $L_c = 5$$\times$$10^{-7}$ are used as the default observer gains. This exact same analysis and design procedure can be used for the modified observer, with the final observer gains selected to be $L_T = -5$$\times$$10^{-3}$ and  $L_c = 1$$\times$$10^{-4}$.

\subsection{Observer performance under various conditions} \label{sec:Performance}
In the previous section, the observer is designed using the default parameter values. This section explores the performance of the designed observer under various conditions.

From the mechanistic understanding of desorption described in Section \ref{sec:Model}, three key parameters that directly influence the desorption dynamics are the frequency factor $A$, activation energy $E_a$, and concentration $c_s$. Hence, we study the performance of the observer for several values of $A$, $E_a$, and $c_{s,0}$ reported in the lyophilization literature.

\begin{figure} [ht!]
\centering
    \includegraphics[scale=.95]{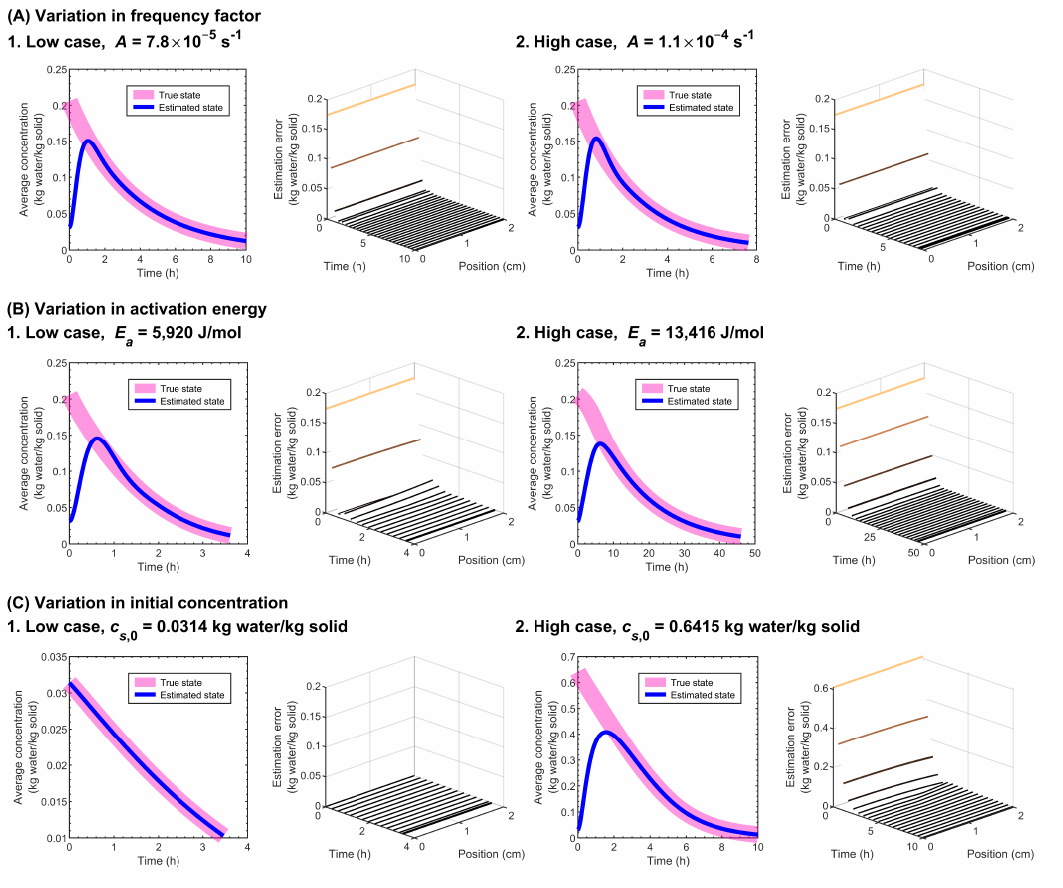}

\vspace{-0.5cm}
    
    \caption{(A) Convergence of the estimated concentration for different frequency factors $A$. The frequency factor is varied between 7.8$\times$10$^{-5}$ s$^{-1}$ \cite{Sheehan_1998_Modeling} and 1.1$\times$10$^{-4}$ s$^{-1}$ \cite{Mascarenhas_1997_FEMmodel} given that the rate constant is independent on temperature. (B) Convergence of the estimated concentration for different activation energies $E_a$. The activation energy is varied between 5,920 J/mol \cite{Fissore_2015_Review}  and 13,146 J/mol \cite{Oddone_2017_Ea}. (C) Convergence of the estimated concentration for different initial concentrations, $c_{s,0}$. The initial condition is varied between 0.0314 kg water/kg solid \cite{Fissore_2018_Review} and 0.6415 kg water/kg solid \cite{Sadikoglu_1997_Modeling}. In all cases, the simulations are terminated when the estimated concentration drops below 0.01 kg water/kg solid. The default observer gains $L_T = -1$$\times$$10^{-6}$ and $L_c = 5$$\times$$10^{-7}$ are used. Other parameters are kept at the default values.} 
    \label{fig:Obs2}      
\end{figure}

The first part of the study focuses on the frequency factor $A$. The estimated state smoothly converges to the true state within 2 h from the time scale of 10 h for the low case (Fig.\ \ref{fig:Obs2}A1). An increase in the frequency factor reduces the time required for secondary drying to less than 8 h, in which the estimated state can converge to the true state within about 90 min (Fig.\ \ref{fig:Obs2}A2). Convergence is achieved for both spatial and average concentration.

The second part of this analysis centers on the activation energy $E_a$. For the low case, the estimated concentration converges to the true value within 1 h from the time scale of about 4 h (Fig.\ \ref{fig:Obs2}B1). The dynamics of the process are relatively fast here compared to other cases, but the observer can still perform very well. For the high case, the convergence is observed at about 6 h from the time scale of about 45 h (Fig.\ \ref{fig:Obs2}B2). The drying time of 45 h is considered extremely slow for secondary drying, but that does not impact the performance of the observer.

The final part of this analysis considers the concentration level $c_{s,0}$. Regarding the low case, the initial estimated concentration is somehow correct, i.e., equal to the true value, so the convergence is immediate (Fig.\ \ref{fig:Obs2}C1). For the high case, the convergence is observed at about 2 h from the time scale of 10 h (Fig.\ \ref{fig:Obs2}C2). The observer can converge quickly even when the initial estimated state is off by more than an order of magnitude; i.e., the initial estimated concentration is 0.0314 kg water/kg solid, whereas the true value is 0.6415 kg water/kg solid.

Results from this study show that the observer is able efficiently and accurately estimate the concentration of bound water for various desorption dynamics considered in the literature. The convergence is achieved for every case study via a single observer design where $L_T = -1$$\times$$10^{-6}$ and $L_c = 5$$\times$$10^{-7}$, indicating that the proposed observer and design strategy are robust.

\subsection{Measurement noise and observer gain scheduling}  \label{sec:Noise}
A state observer always receives some form of measurement, and thus it is important to ensure that the estimated states are insensitive to measurement noise. A high observer gain could give fast convergence but also magnify the noise, resulting in inaccurate estimation of the states. This section discusses effects of measurement noise and demonstrates the corresponding observer design.

\begin{figure} [ht!]
\centering
    \includegraphics[scale=0.95]{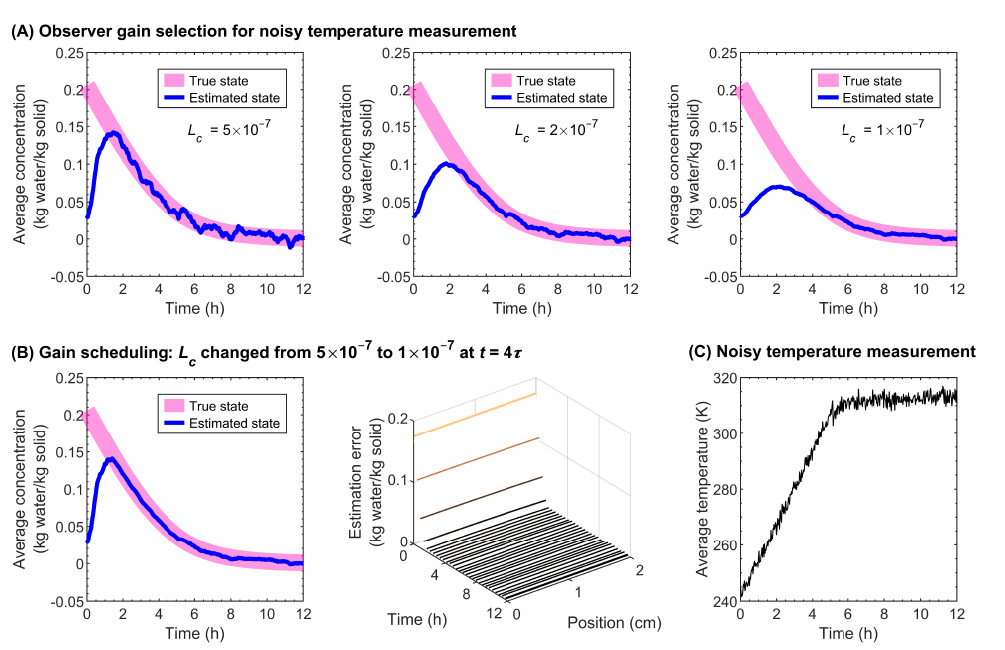}

\vspace{-0.3cm}
    
    \caption{(A) Convergence and oscillation behavior of the estimated concentration at different observer gains $L_c = 5$$\times$$10^{-7}$, $2$$\times$$10^{-7}$, and $1$$\times$$10^{-7}$ while $L_T$ is fixed at $-1$$\times$$10^{-6}$ in all cases. (B) Gain scheduling technique where $L_c$ is set to $5$$\times$$10^{-7}$ for fast convergence at the beginning and then reduced to $1$$\times$$10^{-7}$ at $t = 4\tau$ to reduce the effect of measurement noise. (C) Noisy temperature data obtained from adding independent normally distributed noise of standard deviation given by $3\sigma=5^\circ$C to the true state. Other parameters are kept at their default values. } 
    \label{fig:Obs3}      
\end{figure}

To simulate the noise, independent normally distributed noise of standard deviation given by $3\sigma=5^\circ$C \cite{Srisuma_2023_1DCellThawing} is added to the temperature profile given the default model parameters. The estimated concentration is severely polluted by the measurement noise for the default observer gain $L_c = 5$$\times$$10^{-7}$ (Fig.\ \ref{fig:Obs3}A). Significant oscillation is observed, in which the observer does not give valuable information when the concentration is small (i.e., after about 6 h). To reduce the noise effect, the observer gain $L_c$ is reduced to $2$$\times$$10^{-7}$. In comparison to $L_c = 5$$\times$$10^{-7}$, the convergence is achieved slightly slower, but the oscillation is much weaker. Reducing the gain $L_c$ further to $1$$\times$$10^{-7}$ almost removes the oscillation, but the convergence is also significantly slower.

The above information motivates the use of observer gain scheduling, which is a technique that varies the observer gains with time to ensure satisfactory performance at different operating points, especially for nonlinear systems. In this case, we can initialize the observer with a high gain to achieve fast convergence, and then gradually switch to a low gain after some time to reduce oscillation. Here the switching time is set to $t=4\tau$ following the definition of the convergence time defined in Section \ref{sec:DesignStrategies},
where $\tau$ is the time constant derived in Section \ref{sec:DesignStrategies}. This technique enables fast convergence while having low effects of measurement noise on the state estimate (Fig.\ \ref{fig:Obs3}B). 

This analysis highlights another benefit of the proposed observer in terms of noise filtering. Selecting the observer gain needs to trade off sensitivity to measurement noise with speed of convergence of the state estimates as demonstrated above. For this process, it is acceptable to allow for some small oscillation when the concentration is high (e.g., larger than 0.1) as the value is not affected much. However, oscillation should be minimized when the concentration is low, so that the final concentration of bound water can be accurately estimated to ensure product quality. 

Note that the above noise level (Fig.\ \ref{fig:Obs3}C) is significantly higher than that observed in actual experiments \cite{Srisuma_2023_1DCellThawing}, and so the resulting design has a large safety margin to span the range of noise levels encountered in real temperature sensors. A lower safety margin, with faster convergence to the states, could be achieved by performing the analysis for a noise level set by experimental data for the specific temperature sensor used in the specific equipment.

\subsection{Experimental-based observer design}  \label{sec:RealSystems}
In the simulation-based observer design, the true and estimated states are simulated simultaneously and continuously. In real systems, measurement data are usually sampled and fed to a state observer in a discrete fashion, i.e., with a fixed sampling time. This section applies the proposed observer to estimate the concentration of bound water in four different experiments from the literature, denoted as Cases A (skim milk) \cite{Sadikoglu_1997_Modeling}, B (skim milk) \cite{Liapis_1994_Original}, C (sucrose) \cite{Pisano_2012_ObsSecondary}, and D (mannitol) \cite{Fissore_2015_Review}, with the sampling time of 10 s. Although our model is developed based on lyophilization of unit doses (vials), the experimental data used here include both bulk lyophilization (Cases A and B) and vial lyophilization (Cases C and D) to demonstrate the generalizability of the approach given that radial heat transfer is negligible compared to vertical heat transfer.

Table \ref{Tab:ParametersSpecific} lists the parameter values specific to the four case studies considered here. For Cases A and B, temperature data were not directly reported in the original studies \cite{Liapis_1994_Original,Sadikoglu_1997_Modeling}; however, a mechanistic model and model parameters were given such that spatial temperature data could be simulated. The simulated temperature profile was used as the temperature measurement for our state observer, with the sampling time of 10 s. Additionally, the original residual moisture was reported as the total mass of water, so we normalized the values by the total mass of solid to give the unit of kg water/kg solid. For Cases C and D, the bottom temperatures were reported in the original studies \cite{Pisano_2012_ObsSecondary,Fissore_2015_Review}, but the sampling time is too large (about 30 min), resulting in poor convergence. Hence, based on the reported temperature, we used the model to reconstruct the the temperature profile with the sampling time of 10 s, which was then fed to our observer.

\begin{table}[ht!]
\caption{Specific parameters for the four experimental systems.}
\renewcommand{\arraystretch}{1.2}
\label{Tab:ParametersSpecific}
\centering
\begin{tabular}{| c | c | c |c | c |}
\hline
\textbf{Case}&\textbf{Parameter} & \textbf{Value} & \textbf{Unit} & \textbf{Reference/Note} \\ 
\hline 
\multirow{3}{*}{1} & $E_a$ & 5,000 &  J/mol &  Estimated from data \\
& $A$ &  7.1$\times$10$^{-4}$  &  s$^{-1}$ &   Estimated from data \\
 & $c_{s,0}$ & 0.6415 &  kg water/kg solid & \cite{Sadikoglu_1997_Modeling} \\
\hline
\multirow{4}{*}{B} & $k$ & 0.028 & W/(m$\cdot$K)  &   \cite{Liapis_1994_Original} \\  
& $E_a$ & 5,300 &  J/mol &  Estimated from data \\
& $A$ & 4.5$\times$$10^{-4}$  &  s$^{-1}$ &   Estimated from data \\
& $c_{s,0}$ & 0.1940 &  kg water/kg solid & \cite{Liapis_1994_Original}  \\
 \hline
\multirow{7}{*}{C} & $E_a$ & 37,714 &  J/mol &  \cite{Pisano_2012_ObsSecondary} \\
& $A$ & 277  &  s$^{-1}$ &   \cite{Pisano_2012_ObsSecondary} \\
& $h$ & 7 & W/(m$^2\cdot$K) &   Estimated from data  \\ 
& $T_0$ & 270.38 &  K &  \cite{Pisano_2012_ObsSecondary} \\
& $T_{b,0}$ & 270.38 &  K & Assumed to be $T_0$  \\
& $c_{s,0}$ & 0.0314 &  kg water/kg solid & \cite{Pisano_2012_ObsSecondary} \\
& $r$ & 0.6 &  K/min &  Estimated from data \\ \hline
\multirow{7}{*}{D} & $E_a$ & 5,920 &  J/mol &  \cite{Fissore_2015_Review} \\
& $A$ & 1.2$\times$$10^{-3}$  &  s$^{-1}$ &   Estimated from data \\
& $h$ & 7 & W/(m$^2\cdot$K) &   Estimated from data  \\ 
& $T_0$ & 264.09 &  K &  \cite{Fissore_2015_Review} \\
& $T_{b,0}$ & 264.09 &  K & Assumed to be $T_0$  \\
& $T_{b,\textrm{max}}$ & 312 &  K &  Estimated from data  \\
& $c_{s,0}$ & 0.0603 &  kg water/kg solid & \cite{Fissore_2015_Review} \\
& $r$ & 0.5 &  K/min &  Estimated from data \\
& $H$ & 0.0102 &  m & Calculated from parameters in Ref.\ \cite{Fissore_2015_Review} \\ \hline
\end{tabular}
\renewcommand{\arraystretch}{1}
\end{table}

\begin{figure} [ht!]
\centering
    \includegraphics[scale=1]{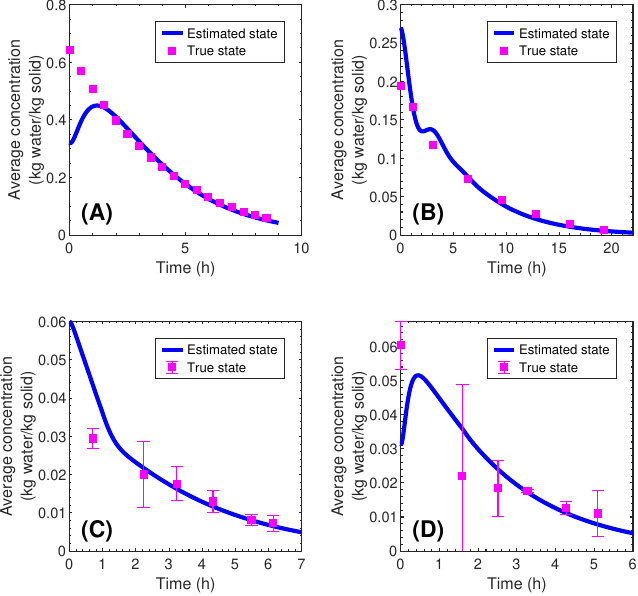}
    \vspace{-0.1cm}
    \caption {Convergence of the estimated concentration when applying the state observer to the real systems in Case A \cite{Sadikoglu_1997_Modeling} (Panel A) and Case B \cite{Liapis_1994_Original} (Panel B). Convergence of the estimated concentration when applying the modified state observer to the real systems in Case C \cite{Pisano_2012_ObsSecondary} (Panel C) and Case D \cite{Fissore_2015_Review} (Panel D). In all case studies, the sampling time of 10 s is used, with the default parameter values and observer gains. The initial estimated concentration is chosen to be lower/higher than the true value by about 20\%--100\% to demonstrate convergence. The initial estimated temperature is set to the measured value.} 
    \label{fig:ObsExp}      
\end{figure}

By using the default observer gains obtained from our simulation-based design, the observer can converge the estimated concentration to the correct value in less than 2 h for all four experiments, showing the robustness of our design (Fig.\ \ref{fig:ObsExp}). In Figs. \ref{fig:ObsExp}AB, two experiments with spatially distributed temperature measurement are considered, and so the state observer is used as usual. In Figs.\ \ref{fig:ObsExp}CD, the only measurement available is the bottom temperature, and so the modified state observer is used instead (see Section \ref{sec:ModObserver}). Both the original and modified observers work perfectly. 

The observer with spatially distributed measurement, e.g., IR cameras, provides more complete information of the product. IR cameras are always located outside a vial, enabling non-contact measurement \cite{Bockstal_2018_IR,Colucci_2020_IR,Harguindeguy_2022_IR}. However, only edge vials could be monitored, as center vials cannot be seen by the camera \cite{Vallan_2023_TempMeasure}. In this case, a model that properly incorporates the effect of thermal radiation can be used to reconstruct the temperature profiles of all vials. Noisy thermal imaging data can be filtered directly with the observer \cite{Srisuma_2023_1DCellThawing} or some filters such as a low-pass filter \cite{Srisuma_2023_1DCellThawing} and Savitzky-Golay ﬁlter \cite{Harguindeguy_2022_IR}. The observer with point measurement, e.g., a thermocouple, would lead to a simpler and less expensive setup, with lower noise and bias \cite{Lietta_2019}. This contact measurement may not be feasible for many pharmaceutical applications where contamination must be minimized. The presence of a thermocouple could also alter local heat transfer near the contact point inside the product, which is not captured by the mechanistic model; see Ref.\ \cite{Vallan_2023_TempMeasure} for detailed discussion on various temperature measurements in lyophilization.

The choice of sampling time, e.g., 10 s used in Fig.\ \ref{fig:ObsExp}, needs to be considered when implementing a state observer to the real system. If the sampling time is too large, the observer could converge slowly or diverge due to insufficient measurement information. Also, the sampling time must be higher than the computation time required for simulating the observer in each time step. Our observer can be simulated in less than a second on a normal laptop, and thus the sampling time of 10 s is more than adequate. The sampling time should also be chosen to be sufficiently small compared to the time scale of a process to ensure that any important dynamics are well captured. For example, the time scale of 10 s is reasonable given the time scale of many hours in lyophilization.

\begin{figure} [ht!]
\centering
    \includegraphics[scale=1]{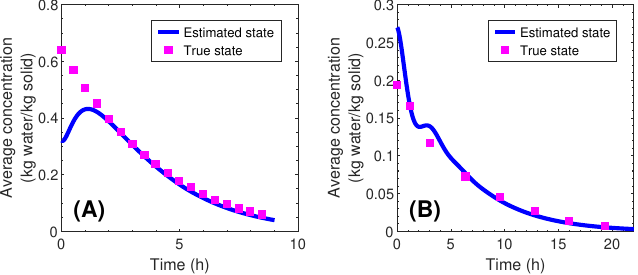}
    \vspace{-0.1cm}   
    \caption {Convergence of the estimated concentration when applying the state observer to Case A \cite{Sadikoglu_1997_Modeling} (Panel A) and Case B \cite{Liapis_1994_Original} (Panel B), with the value of the heat transfer coefficient $h$ for the observer set to be 10\% lower than the correct value for Case A and 10\% higher than the correct value for Case B. The sampling time of 10 s is used, with the default parameter values and observer gains. The initial estimated concentration is chosen to be different from the correct value to demonstrate convergence. The initial estimated temperature is set to the measured value.} 
    \label{fig:ObsFin}      
\end{figure}

The heat transfer dynamics predicted by the model is a function of the heat transfer coefficient $h$, which is estimated from data and so can have some uncertainty. By using the data from Cases A \cite{Sadikoglu_1997_Modeling} and B \cite{Liapis_1994_Original}, we show that our observer converges the estimated states to the correct values even for 10\% error in the value of the heat transfer coefficient (Fig.\ \ref{fig:ObsFin}). The convergence is slightly slower, but the difference is nearly unnoticeable (cf., Figs.\ \ref{fig:ObsExp}AB and \ref{fig:ObsFin}AB). The estimated concentration is almost insensitive to uncertainty in the heat transfer coefficient in these cases.

As shown in Figs.\ \ref{fig:ObsExp} and \ref{fig:ObsFin}, our observer can be applied for a wide range of concentration values. In secondary drying, the observer might be used for different purposes, in which the design and validation should correspond to that application. For example, the observer could be used to monitor the drying process until the moisture content is below a certain threshold. Such applications are common and do not require a highly accurate observer for the entire concentration range because only the threshold is concerned. However, there are some sensitive materials, e.g., biologics, that require precise control within a narrow range of moisture, e.g., 1\%--3\% \cite{Zhang_2021_Biologic}. In such cases, it is crucial to ensure that the observer is very accurate for the desired concentration range, meaning that the mechanistic model and temperature measurement must be highly reliable. Otherwise, using a highly accurate concentration sensor could be a better alternative.

\subsection{Application to microwave lyophilization} \label{sec:Microwave}
All the results presented in the previous sections are based on conventional lyophilization, i.e., $Q_v = 0$. In conventional lyophilization, the product contained in a glass vial is heated from the bottom via a heating shelf during the drying stages \cite{Pisano_2010_control,Fissore_2018_Review}. To accelerate the drying process, various techniques have been studied and developed; one of the most prominent techniques is microwave lyophilization, where microwave irradiation is used to reduce the drying time during primary and secondary drying \cite{Walters_2014_DryingTech,Gitter_2018_Experiment,Gitter_2019_Parameters}.
This section demonstrates the application of our observer for feedback control of microwave lyophilization. 

Control strategies for microwave lyophilization have not been demonstrated well in the literature. It has been shown that continuous microwave heating for water desorption can lead to a huge increase in the product temperature if the microwave power is not well controlled \cite{Witkiewicz_2014_MFDcontrol}. Overheating is not desirable in lyophilization as it could lead to some serious issues such as product collapse and liquid-phase formation \cite{Fissore_2018_Review}. As such, the main goal of our feedback control system is to control the microwave power $Q_v$ to ensure that the product temperature does not exceed its upper limit. Here we define $Q_v$ by using a simple proportional control relation:
\begin{equation}
    Q_v(t) = K(T_\text{up}-T_\text{max}(t)),    
\end{equation}
where $K$ is the controller gain, $T_\text{up}$ is the upper temperature limit, and $T_\text{max}$ is the maximum product temperature given the temperature distribution $T(z,t)$. Here we specify $K = 1,000$ W/(m$^3\cdot$K) and $T_\text{up} = T_{b,\textrm{max}}$. Other parameters are based on their default values.

\begin{figure} [ht!]
\centering
    \includegraphics[scale=1]{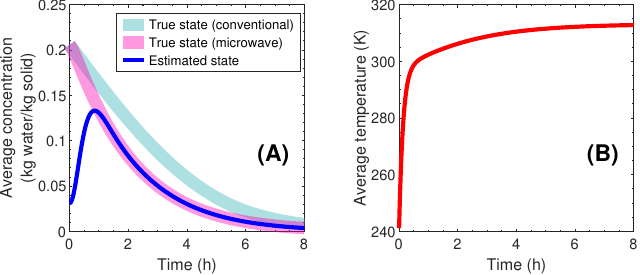}

    \vspace{-.1cm}
    
    \caption {(A) Convergence of the estimated concentration for the observer-based feedback control system where the microwave power is controlled to prevent product overheating, compared to conventional lyophilization. (B) Average product temperature under controlled microwave heating. The default observer gains $L_T = -1$$\times$$10^{-6}$ and $L_c = 5$$\times$$10^{-7}$ are used. Other parameters are kept at the default values.} 
    \label{fig:ObsMFD}      
\end{figure}

By using the default observer gains, the observer can accurately estimate the concentration of bound water despite the presence of microwave irradiation and feedback control (Fig.\ \ref{fig:ObsMFD}A). In comparison to conventional lyophilization, the drying time is shortened by about 2 h. With the proposed proportional control, the product temperature is well controlled below the maximum temperature of about 313.15 K (Fig.\ \ref{fig:ObsMFD}B).

\section{Conclusion} \label{sec:Conclusion}
A novel approach for the real-time estimation of the residual moisture during secondary drying in lyophilization is presented. The technique relies on a state observer, in particular a Luenberger observer, which uses the information from mechanistic understanding of the process and temperature measurement to predict the concentration of bound water. Our observer can accurately estimate the amount of bound water for various desorption dynamics, noisy data, and real experiments. Nearly all the case studies presented in this work, except for the noise and gain-scheduling parts, are achieved by a single observer design, indicating high robustness of the observer.

The proposed framework is designed to be simple and practical for implementation. The observer can be simulated in real time, with the computation time of less than a second on a normal laptop. No concentration measurement is required; only temperature measurement is necessary. As temperature measurement is straightforward and commonly required in every step of lyophilization, the technique can be employed with a very simple setup and operation compared to any other methods. The approach is presented systematically, with detailed derivation and mathematical analysis, and so it can be easily extended to desorption-based processes other than lyophilization. This extension can be done by rewriting a mechanistic model for the new system and redesigning a state observer using the procedure described in this article.



\clearpage


\pagebreak

\section*{Data Availability}  \label{sec:Code}
Software and data used in this work are available at \url{https://github.com/PrakitrSrisuma/lyo-observer-secondary}.

\section*{Acknowledgements} 
This research was supported by the U.S. Food and Drug Administration under the FDA BAA-22-00123 program, Award Number 75F40122C00200.

\section*{Appendices}
\appendix
\renewcommand\thefigure{\thesection.\arabic{figure}} 
\renewcommand\theequation{\thesection.\arabic{equation}}

\setcounter{figure}{0} 
\setcounter{equation}{0} 
\section{Numerical methods} \label{app:Model_FVM}
\begin{figure} [ht!]
\centering
    \includegraphics[scale=0.6]{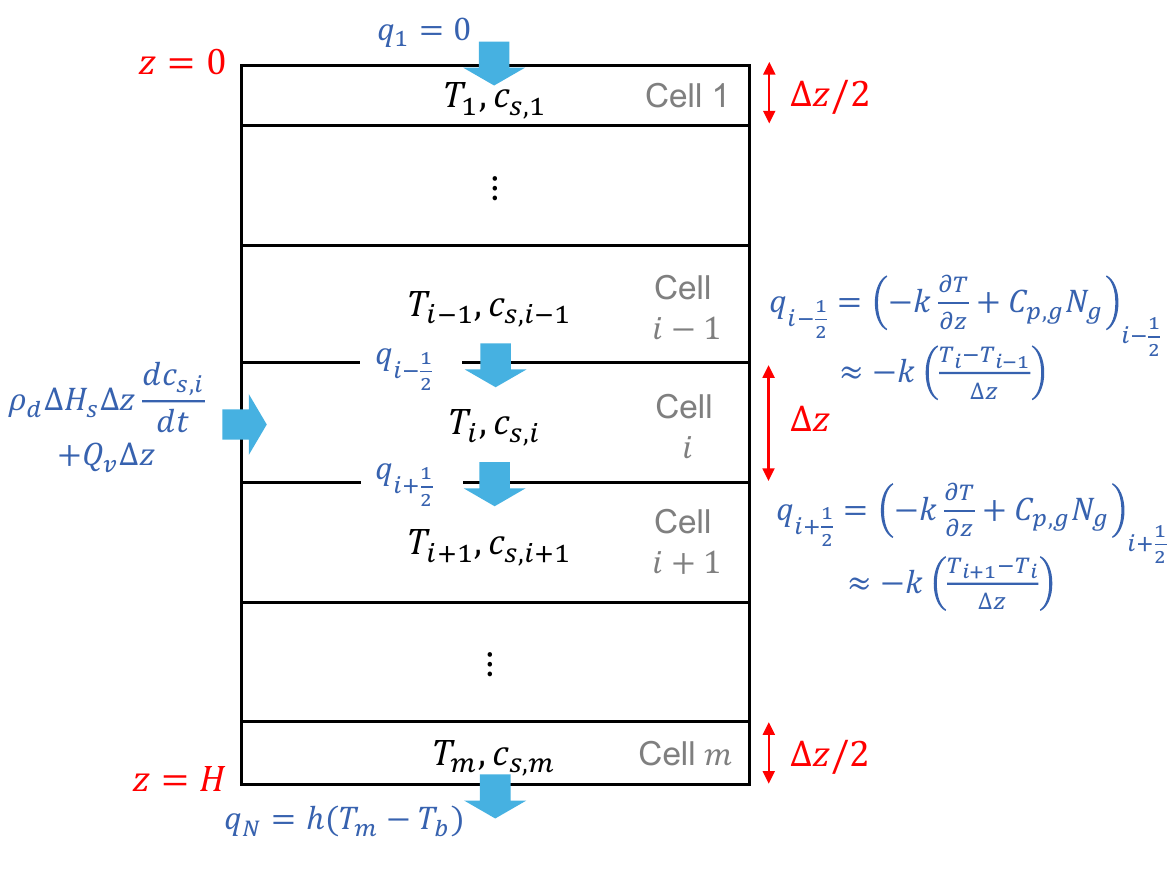}
    \caption {Finite volume discretization of the model equations. The variables $T_i$ and $c_{s,i}$ denote the temperature and bound water concentration of cell $i$.} 
    \label{fig:FVM}      
\end{figure}

The mechanistic model is presented in Section \ref{sec:Model}. To simulate the model, various numerical methods can be employed. Since the model equations contain the flux terms and boundary conditions, the finite volume method (FVM) is selected. The spatial domain is discretized into $m$ cells of volume $A_c\Delta z$ as shown in Fig.\ \ref{fig:FVM}, where $A_c$ is the cross sectional area and $\Delta z$ is given by
\begin{equation} \label{eq:dx}
    \Delta z = \frac{H}{m-1}.
\end{equation}

Considering each cell as a control volume, the energy balances are
\begin{align}
&A_c \frac{\Delta z}{2} \rho C_p \frac{dT_1}{dt}  = q_1A_c - q_{{3}/{2}}A_c + A_c\frac{\Delta z}{2}\rho_d\Delta H_s\dfrac{dc_{s,1}}{dt} + A_c\frac{\Delta z}{2} Q_v, \\
&A_c \Delta z \rho C_p \frac{dT_i}{dt}  = q_{i-{1}/{2}}A_c - q_{i+{1}/{2}}A_c + A_c\Delta z\rho_d\Delta H_s \dfrac{dc_{s,i}}{dt} + A_c\Delta z Q_v,  \qquad \textrm{for} \ i = 2,\dots,m-1, \\
& A_c \frac{\Delta z}{2} \rho C_p \frac{dT_m}{dt}  =  q_{m-{1}/{2}}A_c - q_{m}A_c + A_c\frac{\Delta z}{2}\rho_d\Delta H_s\dfrac{dc_{s,m}}{dt} + A_c\frac{\Delta z}{2} Q_v.
\end{align}
As a result, the final discretized equations are
\begin{align}
    & \frac{dT_1}{dt}  = \frac{2k}{\rho C_p}\!\left(\frac{T_2-T_1}{\Delta z^2}\right)\! + \frac{\rho_d\Delta H_s}{\rho C_p}\!\left(\dfrac{dc_{s,1}}{dt}\right)\! + \frac{Q_v}{\rho C_p}, \label{eq:energy_discretized1}\\
    & \frac{dT_i}{dt}  = \frac{k}{\rho C_p}\!\left(\frac{T_{i+1}-2T_i+T_{i-1}}{\Delta z^2}\right)\!+ \frac{\rho_d\Delta H_s}{\rho C_p}\!\left(\dfrac{dc_{s,i}}{dt}\right)\! + \frac{Q_v}{\rho C_p}, \qquad   \textrm{for} \ i = 2,\dots,m-1, \label{eq:energy_discretizedi} \\
   & \frac{dT_m}{dt} = -\frac{2k}{\rho C_p}\!\left(\frac{T_m-T_{m-1}}{\Delta z^2}\right)\! - \frac{2h}{\rho C_p \Delta z}(T_m-T_b) + \frac{\rho_d\Delta H_s}{\rho C_p}\!\left(\dfrac{dc_{s,m}}{dt}\right) \!+ \frac{Q_v}{\rho C_p}. \label{eq:energy_discretizedN}
\end{align}
For the desorption process, the discretized equations are
\begin{equation} \label{eq:desorption_discretized}
    \dfrac{dc_{s,i}}{dt} = Ae^{-E_a/RT_i} c_{s,i}, \qquad   \textrm{for} \ i = 1,2,\dots,m.
\end{equation}
Here we use the finite volume method to transform the original partial differential equations (PDEs) into a system of ordinary differential equations (ODEs). This technique is known as the method of lines \cite{Schiesser_1991_MOL}. The system of ODEs given by \eqref{eq:energy_discretized1}--\eqref{eq:desorption_discretized} can be integrated by commercial ODE solvers; MATLAB's \texttt{ode15s} is used in this work.

Finally, define the state $\mathbf{x}$ and manipulated variable $\mathbf{u}$ as
\begin{gather}
    \mathbf{x} = \!\begin{bmatrix}
     \mathbf{T}  \\
     \mathbf{c_s}
    \end{bmatrix}\!, \label{eq:state2}\\
    \mathbf{u} =\! \begin{bmatrix}
         T_b  \\
         Q_v
    \end{bmatrix}\!,
\end{gather}
where $\mathbf{T} \in \mathbb R^m$ collects the product temperatures ($T_1,\dots,T_n$) and  $\mathbf{c_s} \in \mathbb R^m$ collects the bound water concentrations ($c_{s,1},\dots,c_{s,n}$). The heating shelf temperature $T_b$ and microwave power $Q_v$ can be manipulated to control the dynamic of secondary drying. The model equations are nonlinear in $\mathbf{x}$ and linear in $\mathbf{u}$. Consequently, \eqref{eq:energy_discretized1}--\eqref{eq:desorption_discretized} can be written in the vector form
\begin{equation} \label{eq:FinalODEs_1}
    \dfrac{d\mathbf{x}}{dt} = \mathbf{F(x)} + \mathbf{Bu},
\end{equation}
where $\mathbf F \in \mathbb R^{2m}$ is the corresponding nonlinear function and $\mathbf B \in \mathbb R^{2m\times2}$ is the matrix.

\setcounter{figure}{0} 
\setcounter{equation}{0} 
\section{Time constant approximation}  \label{app:TimeConstant}
Consider a linear system of ordinary differential equations (ODEs)
\begin{equation} \label{eq:linearODE}
    \frac{d\mathbf{x}}{dt} = \mathbf{Mx},
\end{equation}
where $\mathbf{x}$ is the state vector and $\mathbf{M}$ is the matrix of coefficients. Note that here $\mathbf{x}$ is a general state vector, which is not the same definition as \eqref{eq:state2}. The matrix $\mathbf{M}$ can be factorized as
\begin{equation} \label{eq:factorize}
    \mathbf{M} = \mathbf{U \Lambda U}^{-1},
\end{equation}
where $\mathbf{\Lambda}$ is a diagonal matrix containing the eigenvalues of $\mathbf{M}$ and $\mathbf{U}$ is the matrix of the corresponding eigenvectors.\footnote{This analysis assumes that the matrix $\mathbf{M}$ is diagonalizable, which is true for the relevant matrix in this article. A more general theoretical analysis that gives similar results is based on the Jordan form for $\mathbf{M}$, and is provided in textbooks on dynamical systems analysis or state-space control.} Rearranging \eqref{eq:linearODE} with \eqref{eq:factorize} gives
\begin{equation} \label{eq:linearODE2}
    \frac{d}{dt} \!\left(\mathbf{U}^{-1}\mathbf{x}\right)\! = \mathbf{\Lambda} \!\left(\mathbf{U}^{-1}\mathbf{x}\right).
\end{equation}
For a new set of variables defined by
\begin{equation} \label{eq:new}
    \mathbf{y} = \mathbf{U}^{-1}\mathbf{x},
\end{equation}
\eqref{eq:new} can be rewritten as
\begin{equation} \label{eq:linearODE3}
    \frac{d\mathbf{y}}{dt} = \mathbf{\Lambda y}.
\end{equation}
Here the system of equations is completely decoupled. For example, consider the case where $\mathbf{x}\in \mathbb R^{2}$, the corresponding vectors and matrices are
\begin{gather}
    \mathbf{x} = \!\begin{bmatrix}
                    x_1 \\
                    x_2
                \end{bmatrix}, \\
   \mathbf{y} = \!\begin{bmatrix}
                    y_1 \\
                    y_2
                \end{bmatrix}, \\
   \mathbf{\Lambda} = \!\begin{bmatrix}
                  \lambda_1  & 0\\
                    0 & \lambda_2
                \end{bmatrix}, \\
   \mathbf{U} = \!\begin{bmatrix}
                  u_{1,1}  & u_{1,2}\\
                  u_{2,1} & u_{2,2}
                \end{bmatrix}.    
\end{gather}
Consequently, \eqref{eq:linearODE3} becomes
\begin{equation} \label{eq:linearODE4}
    \frac{d}{dt}\! \begin{bmatrix}
                    y_1 \\
                    y_2
                \end{bmatrix}\!
                = \!\begin{bmatrix}
                    \lambda_1  & 0\\
                    0 & \lambda_2
                \end{bmatrix}\!\begin{bmatrix}
                    y_1 \\
                    y_2
                \end{bmatrix},
\end{equation}
where the analytical solution is
\begin{equation} \label{eq:sol1}
 \begin{bmatrix}
        y_1 \\[.1cm] 
        y_2
    \end{bmatrix}
    \!=\! \begin{bmatrix}
        y_{1,0}\: \mathrm{exp}(\lambda_1t) \\[.1cm] 
        y_{2,0}\: \mathrm{exp}(\lambda_2t)
    \end{bmatrix}.
\end{equation}
Multiplying both sides of \eqref{eq:sol1} with the matrix $\mathbf{U}$ results in
\begin{equation} \label{eq:sol2}
 \begin{bmatrix}
        x_1 \\[.1cm] 
        x_2
    \end{bmatrix}
   \! = \mathbf{U}\! \begin{bmatrix}
        y_{1,0}\: \mathrm{exp}(\lambda_1t) \\[.1cm] 
        y_{2,0}\: \mathrm{exp}(\lambda_2t)
    \end{bmatrix}.
\end{equation}
Representing the inverse of $U$ as
\begin{equation}
   \mathbf{U}^{-1} = \!\begin{bmatrix}
                  v_{1,1}  & v_{1,2}\\
                  v_{2,1} & v_{2,2}
                \end{bmatrix}, 
\end{equation}
the initial condition of $\mathbf{y}$ can be written as
\begin{equation} \label{eq:ini}
     \begin{bmatrix}
        y_{1,0} \\[.1cm] 
        y_{2,0}
    \end{bmatrix}
    \!=\! \begin{bmatrix}
        v_{1,1}x_{1,0} + v_{1,2}x_{2,0} \\[.1cm] 
        v_{2,1}x_{1,0} + v_{2,2}x_{2,0}
    \end{bmatrix}.
\end{equation}
Insertion of \eqref{eq:ini} into \eqref{eq:sol2} gives the analytical solution
\begin{equation} \label{eq:sol3}
     \begin{bmatrix}
        x_1 \\[.1cm] 
        x_2
    \end{bmatrix}
   \! =\! \begin{bmatrix}
        u_{1,1}(v_{1,1}x_{1,0} + v_{1,2}x_{2,0})\: \mathrm{exp}(\lambda_1t) + u_{1,2}(v_{2,1}x_{1,0} + v_{2,2}x_{2,0}) \: \mathrm{exp}(\lambda_2t)  \\[.1cm] 
         u_{2,1}(v_{1,1}x_{1,0} + v_{1,2}x_{2,0})\: \mathrm{exp}(\lambda_1t) + u_{2,2}(v_{2,1}x_{1,0} + v_{2,2}x_{2,0}) \: \mathrm{exp}(\lambda_2t)
    \end{bmatrix}.
\end{equation}
As derived in Appendix \ref{app:Model_FVM}, we consider $\mathbf{x}\in \mathbb R^{2m}$. In this case, the analytical solution is
\begin{equation} \label{eq:sol4}
    x_i = \sum_{p=1}^{2m} \nu_{i,p} \: \mathrm{exp}(\lambda_pt),
\end{equation}
with the coefficient 
\begin{equation} \label{eq:sol5}
    \nu_{i,p} = u_{i,p} \!\left(\sum_{q=1}^{2m} v_{p,q}x_{q,0}\right).
\end{equation}
The coefficient $\nu_{i,p}$ represents the contribution of each eigenvalue $\lambda_p$ to the solution. By plotting the real part of $\nu_{i,p}$, the contribution of each eigenvalue can be quantified. 

\begin{figure} [ht!]
\centering
    \includegraphics[scale=1]{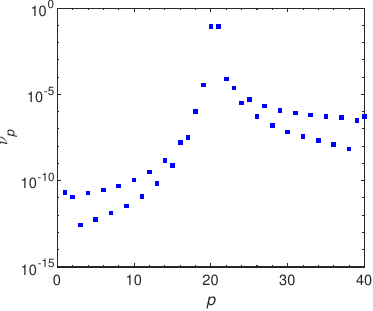}
    \caption {Contribution of each eigenvalue. Each coefficient $\nu_p$ shows the contribution of each eigenvalue $\lambda_p$, with $m = 20$. The default parameters and observer gains are used in this plot. } 
    \label{fig:eigenplot}      
\end{figure}

For the analysis of the observer, the vector $\mathbf{x}$ is replaced with the estimation error, and the matrix $\mathbf{M}$ is replaced by the matrix $\mathbf{F'_{ref}}+\mathbf{LC}$ derived in Section \ref{sec:DesignStrategies}, and thus $\lambda_1, \lambda_2, \dots, \lambda_{2m}$ are the eigenvalues of  $\mathbf{F'_{ref}}+\mathbf{LC}$. Our state of interest is the concentration, which corresponds to $x_{m+1},x_{m+2},\dots,x_{2m}$. The average coefficient for the concentration part $i=m+1,\dots,2m$ is
\begin{equation} \label{eq:sol6}
    \nu_{p} = \frac{1}{m} \sum_{i=m+1}^{2m} \nu_{i,p}.
\end{equation}
By using the default parameter values and observer gains given in Table \ref{Tab:Parameters}, it is shown that $\lambda_m$ and $\lambda_{m+1}$ ($m=20$) have the largest contribution because the coefficients $\nu_m$ and $\nu_{m+1}$ are much larger than the rest by more than three orders of magnitude (Fig.\ \ref{fig:eigenplot}). Therefore, the error dynamics are governed by these two eigenvalues. Since $\lambda_{m+1}$ is smaller than  $\lambda_m$, it suggests the approximation
\begin{equation} \label{eq:timeconstant_new}
    \tau = \frac{1}{\left|\mathrm{Re}(\lambda_{m+1})\right|},
\end{equation}
as described in Section \ref{sec:DesignStrategies}.

\clearpage


\bibliographystyle{elsarticle-num}
\end{document}